\documentclass{emulateapj}
\usepackage{lscape} 
\usepackage{graphics}

\newcommand{\kms}{km\ s$^{-1}$}

\newcommand{\Ha}{H$\alpha$\ }

\newenvironment{inlinefigure}{
\def\@captype{figure}
\noindent\begin{minipage}{0.999\linewidth}\begin{center}}
{\end{center}\end{minipage}\smallskip}

\slugcomment{To appear in ApJL}

\shorttitle{Barred Spirals at z $>$ 0.7} \shortauthors{Sheth et al.}

\begin{document}

\title{Barred Spiral Galaxies at z $>$ 0.7: NICMOS Deep Field
Observations}

\author{Kartik Sheth\altaffilmark{1,2}, Michael W. Regan \altaffilmark{3}, Nicholas Z. Scoville\altaffilmark{1} \& Linda E. Strubbe\altaffilmark{1}}

\altaffiltext{1}{Astronomy Department, California Institute of Technology, Pasadena, CA 91125}
\altaffiltext{2}{kartik@astro.caltech.edu}
\altaffiltext{3}{Space Telescope Science Institute, 3700 San Martin Drive, Baltimore, MD 21218}


\begin{abstract}
Previous optical studies found an unexpected deficit of bars at z $>$
0.7.  To investigate the effects of bandshifting, we have studied the
fraction of barred spirals in the NICMOS Deep Field North.  At z $>$
0.7 we find at least four barred spirals, doubling the number
previously detected.  The number of barred galaxies is small because these
(and previous) data lack adequate spatial resolution.  A typical 5 kpc
bar at z $>$ 0.7 is only marginally detectable for WFPC2 at
0.8$\micron$; the NICMOS data have even lower resolution and can only
find the largest bars.  The average size of the four bars seen at z
$>$ 0.7 is 12 kpc.  The fraction of such large bars (4/95) is higher
than that seen in nearby spirals (1/44); all known selection effects
suggest that the observed fraction is a lower limit.  However,
important caveats such as small numbers and difficulties in defining
comparable samples at high and low redshifts should be noted.  We
conclude that there is no significant evidence for a decrease in the
fraction of barred spirals beyond z$\sim$0.7.

\end{abstract}


\section{Introduction}\label{intro}

When did the first galaxy disks form? How did they evolve?  These
fundamental questions may be addressed by studying barred spiral
galaxies.  Over the last three decades, various studies have shown
that any massive, rotationally-supported, and dynamically cold disk
should be unstable to bar formation (e.g., \citealt{ostriker73,
sellwood93} and references therein); recent studies indicate that even
in massive dark matter halos, large bars may form easily
\citep{athanassoula02}.  Transient bars may also form in mergers
(e.g., \citealt{mihos94}).  Bars have a significant impact on the
subsequent evolution of the disk.  Bars transport massive amounts of
gas to the centers \citep{sakamoto99,sheth03}, ignite circumnuclear
starbursts (e.g., \citealt{ho97} and references therein), and reduce
the chemical abundance gradient \citep{martin94,zaritsky94}. Various
models have indicated that the bar instability may be accompanied by
bulge formation \citep{norman96, friedli93, stanek03}; bulges may 
also evolve by gas transported inwards by the bar \citep{friedli95}.  Thus
the presence and cosmological evolution of barred spirals is integral
to our understanding of galaxy formation and evolution.

Before the Hubble Space Telescope\footnote{The results presented here
were based on observations with the NASA/ESA Hubble Space Telescope
obtained at the Space Telescope Science Institute, which is operated
by the Association of Universities for Research in Astronomy,
Incorporated, under NASA contract NAS5-26555.} (HST), morphological
studies of nascent\footnote{We assume a cosmology with
H$_0$=70,$\Omega_M$=0.3, $\Omega_{\Lambda}$=0.7 throughout this {\sl
Letter}.  In this cosmology z=1 corresponds to 7.7 Gyr ago.} galaxy
disks were difficult.  At high redshifts one expected barred galaxies
to be fairly common because hierarchical clustering models indicated
dynamically colder disks (e.g., \citealt{navarro95}, but see their
caveats).  Also mergers were more frequent in the past
(\citealt{baugh96,ferguson00} and references therein).
Therefore, it was a stunning surprise when the pioneering studies of
galaxy morphology with the Hubble Deep Fields (HDF) found a paucity of
barred spirals beyond z$\sim$0.5--0.7 \citep{vanden96,abraham99}.
Analysis with the Caltech Faint Galaxy Redshift Survey was also
consistent with these conclusions \citep{vanden00}. However,
\citet{bunker99} showed one example where a bar was clearly evident in
a NICMOS (1.6$\micron$) image, but not in the optical HDF data.  Could
the decline in bars be an observational bias?  This question is the
focal point of this {\sl Letter}.

We use NICMOS HDFN images to investigate the bar fraction at z $>$
0.7.  In addition to the expected effect of bandshifting, we show how
limited spatial resolution can limit the identification of bars.  We
find that the fraction of large bars at z $>$ 0.7 may be higher than
that seen in the local Universe and discuss the implications in \S
\ref{impl}.

\subsection{Bandshifting: The Need for NICMOS}\label{band}
For studying barred spirals at high redshifts (z $>$ 0.7), the NICMOS
HDF dataset is ideal because it provides rest-frame V through I-band
images of galaxies.  The better visibility of bars at longer
wavelengths is a well-known effect; unlike spiral arms which are
dominated by blue light from massive, young stars, bars are primarily
composed of old, red, low-mass stars, best traced in the infrared.
Another important consideration is the non-uniform dust obscuration
and peculiar star formation morphology in bars.  Early type barred
spirals are often completely devoid of star formation activity in the
bar, making the bar invisible in blue and ultra-violet bands; only in
the I-band images does a bar become visible.  An example of this is
shown in Figure \ref{fig4303}.  These effects become important with
increasing z because at z=0.7, the reddest WFPC2 filter, F814W,
observes rest-frame $\lambda_o \sim$5000\AA\ making it difficult to
detect a barred galaxy.

\begin{inlinefigure}
\begin{center}
\resizebox{\textwidth}{!}{\includegraphics{f1.eps}}
\end{center}
\figcaption{Left panel: UV appearance of NGC 4303, simulated using a
continuum-subtracted \Ha image.  Note how the bar is completely
invisible.  Middle panel: Blue-band image.  The bar is faint and
difficult to identify.  Right panel: I-band image. Only now does the
bar become visible.  In a companion paper, \citet{strubbe03} simulate
the appearance of nearby galaxies in the HDF confirming that high
redshift (z $>$ 0.7) barred spirals are difficult to identify using
optical filters, consistent with studies by \citet{vanden02a}, and
\citet{whyte02}.}
\label{fig4303}
\end{inlinefigure}

\section{Data Analysis \& Results}
\subsection{Identifying Barred Galaxies in the Deep Field}

There are 904 galaxies with photometric or spectroscopic redshifts in
the HDFN\footnote{We obtained the redshifts and other basic data for
the Deep Field from the NASA/IPAC Extragalactic Database (NED) Level 5
page and the pages designed by S. Gwyn. This research made use of
the NED which is operated by the Jet Propulsion Laboratory, California
Institute of Technology, under contract with the National Aeronautics
and Space Administration.}.  Of these, 206 are at z $<$ 0.7, 226 are
at 0.7 $<$ z $<$ 1.1, and the remaining 472 are at z $>$ 1.1.  We made
postage stamp images of each of these galaxies in the optical and
near-infrared HDF data, examined each galaxy individually in the V
(F606W), I (F814W) and H (F160W) bands, and identified a candidate
sample of 136 galaxies (41 at z $<$ 0.7, 50 at 0.7 $<$ z $<$ 1.1) with
resolved disk morphology.

Bars are characterized by isophotes that have a relatively constant
position angle and a monotonically increasing ellipticity.  At the end
of the bar the ellipticity drops sharply, and the position angle
changes as the isophotes belonging to the underlying disk are fitted.
For each of the 136 galaxies, we used the standard ELLIPSE routine in
IRAF to fit the isophotes with ellipses to identify bars.  This
technique, described by \citet{regan97}, is perhaps the most widely
used technique for identifying bars \citep{martini01,laine02,sheth03}.
However, bars may be missed if the galaxy is highly inclined, if the
bar position angle is the same as the galactic disk, if the underlying
galactic disk is too faint to be adequately imaged, or if the data
have inadequate resolution to resolve bars.  All of these effects {\sl
reduce} the fraction of observed barred spirals.  We emphasize that
the ellipse fitting technique is unlikely to overestimate the fraction
of barred galaxies.

At z $<$ 0.7, we identify five barred spirals, and two candidate
barred spirals, consistent with the prior analysis of the HDFN by
\citet{abraham99}, who also found seven barred galaxies at z $<$ 0.7.
This agreement is not surprising because in this redshift range the
bandshifting effects are not severe.

\begin{inlinefigure}
\begin{center}
\resizebox{\textwidth}{!}{\includegraphics{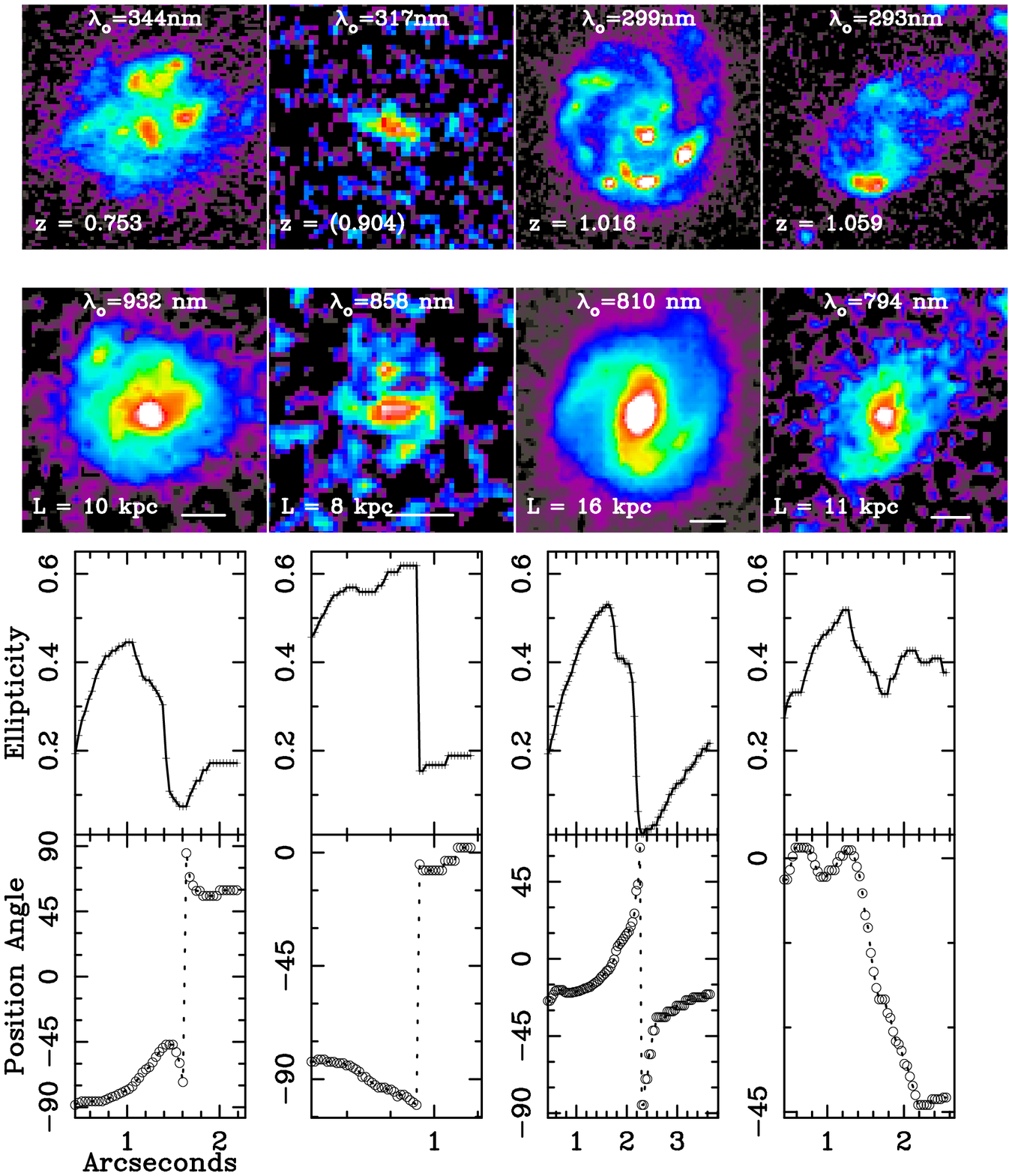}}
\end{center}
\figcaption{Barred spirals at z $>$ 0.7, arranged by redshift.  The top
row shows the optical (F606W, V-band) WFPC2 images and the second row
shows the near-infrared (1.6$\micron$, H-band) NICMOS images.
0.5$\arcsec$ scale is shown with a horizontal segment in the lower
right of each panel.  The rest-frame wavelength for each galaxy is
listed inside the top of each panel.}
\label{bargal}
\end{inlinefigure}

In the redshift range, z $>$ 0.7 we identify four barred spirals and
five candidate barred spirals, including two possible candidates at
z=1.66 and z=2.37.  The four barred spirals are shown in Figure
\ref{bargal}; their properties are listed in Table \ref{bartab}.  For
comparison, in the previous WFPC2 HDFN studies, \cite{vanden96} found
no barred spirals, and \cite{abraham99} found only two barred galaxies
beyond z$\sim$0.5 (see Figure 4 in \citealt{abraham99}).  We note that
our criteria are even more conservative than those used by
\citet{abraham99} who identified bars by calculating a bar axial ratio
parameter from only one outer and one inner isophote, and the
difference in the position angles of the two isophotes; 
we use ellipse fitting over the
entire image to determine the existence of the bar.  In fact, if we
apply the \citet{abraham99} magnitude cutoff of I(AB)=23.7, the
total number of disk-like galaxies drops to only 31 galaxies at
z $>$ 0.7.  Among these we would identify three barred spirals.  As we
discuss later, this fraction of barred spirals is consistent with the
fraction of bars not declining at z $>$ 0.7.  It is important to note,
however, that we are studying galaxies, not in the I-band, but in the
H-band, where typical disks are $\sim$2 magnitudes brighter.

Nevertheless, we detect only a few barred spirals.  Does this reflect
a true decline in barred spirals beyond z $>$ 0.7, as has been suggested
previously?

\subsection{Spatial Resolution \& the Visibility of Bars}\label{resol}

In Figure \ref{angsize}, we show the apparent angular size of various
galactic structures as a function of redshift.  Overlaid are detection
limits for various telescopes adopting a five PSF detection threshold.
A typical bar in the nearby Universe has a size of 5 kpc
\citep{sheth03}. At z $>$ 0.7, the 0.8$\micron$ WFPC2 data is only
marginally capable of detecting such bars.  Combined with the
bandshifting effect, it is thus not surprising that the optical HDF
data showed a decline in the bar fraction at z $>$ 0.7.

\begin{inlinefigure}
\begin{center}
\resizebox{\textwidth}{!}{\includegraphics{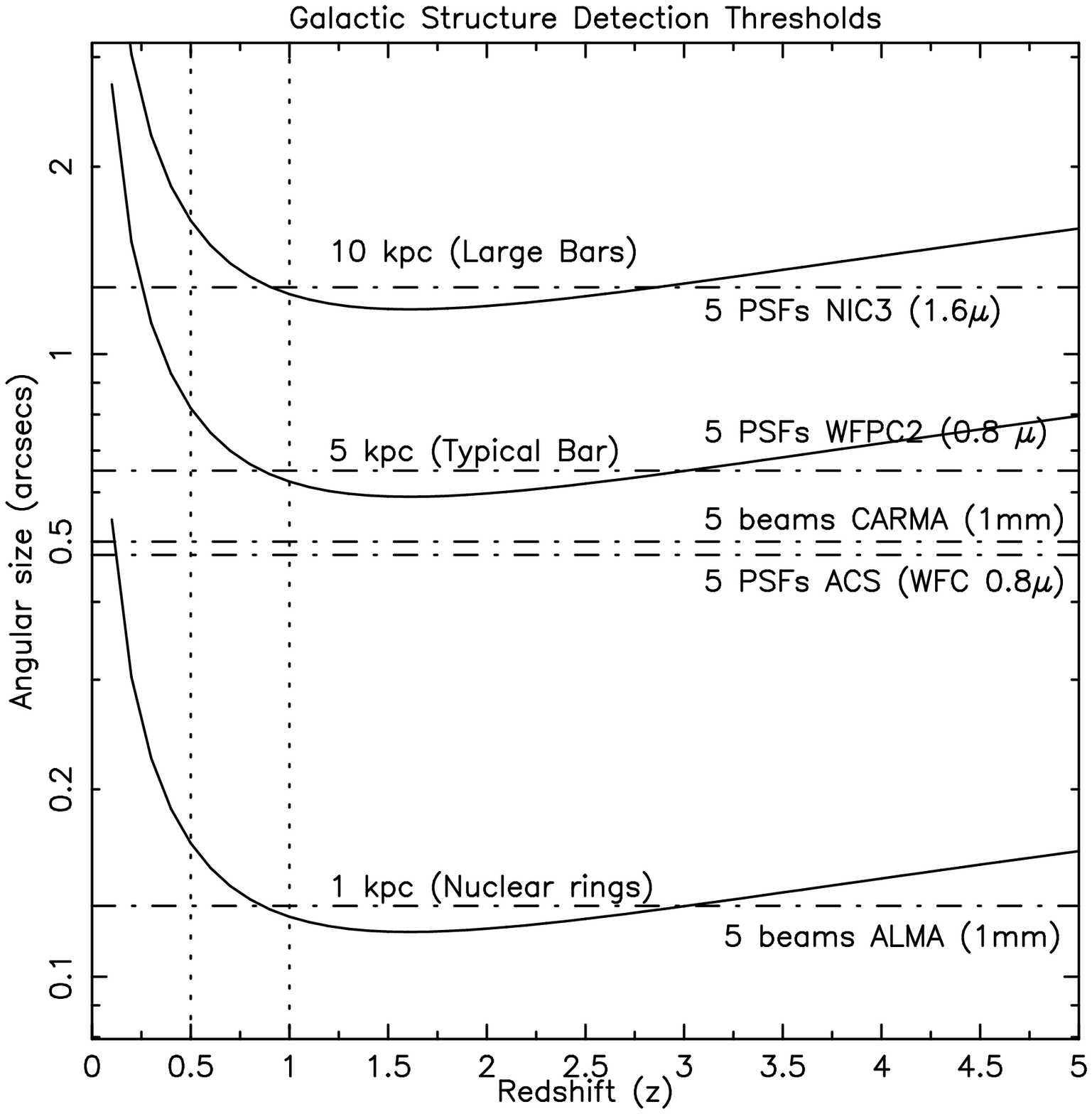}}
\end{center}
\figcaption{The detection threshold of various galactic structures as a
function of redshift for different telescopes and instruments is
shown.  The horizontal dotted-dashed lines are an arbitrary 5 PSF or 5
beam limit.  Note that at 0.8$\micron$, even the WFPC2 data is only
marginally capable of detecting a typical 5 kpc bar at z $>$ 0.7; ACS
z-band is only slightly better.  The NICMOS data can only detect the
largest bars at z $>$ 0.5.  Also shown are capabilities of two new
millimeter arrays, CARMA and ALMA which will be ideal for probing the
gas kinematics in high redshift systems.}
\label{angsize}
\end{inlinefigure}

The NICMOS data are not affected by bandshifting until z $>$ 2--3;
however, these data have even lower resolution than the WFPC2
data due to the longer
wavelengths and larger pixel size.  The NICMOS data can therefore only
detect the largest bars.  The average size of the four bars identified
at z $>$ 0.7 is 12 kpc.

\section{Discussion}
\subsection{Evaluating the Bar Fraction at High and Low Redshifts}

How common are large bars in the local Universe?  In a representative
survey of nearby spiral galaxies (SONG, \citealt{regan01}), there is
only one bar out of 44 spirals with a size larger than 12 kpc
\citep{sheth03}.  This sample was chosen using the following criteria:
all spiral galaxies with M$_B < $ --21.3, $i <$70$^o$, $\delta >$
20$^o$, and V$_{HEL} < $ 2000 \kms.  In nearby spirals, the bar size
is correlated with bulge size \citep{ath80, elmelm85}, but little else
is known about the properties of galaxies that host the largest bars.

To study the bar fraction at z $>$ 0.7, one would ideally like to define
a comparable sample.  However, galaxies at high redshift have
properties that are not yet fully understood.  For instance, the
fraction of irregular or anomalous objects increases with redshifts
(e.g., \citealt{griffiths94, abraham96}).  In a study of seven high
redshift spirals, \citet{quillen98} find M/L ratios that are
comparable to local galaxies, but their analysis indicates that
a fraction of the light must originate from a thick disk/bulge, or
a significant amount of dark matter must be present in the luminous
regions.  For a given luminosity, disks appear to be more compact
\citep{griffiths94}; yet, it is unclear whether massive disks were in
place at z$\sim$1, evolving only passively in luminosity
\citep{lilly96}, or whether significant evolution has occurred
\citep{shude98}.

Keeping these caveats in mind, we consider the fraction of bars in
galaxies with disk-like morphologies (95 galaxies at z $>$ 0.7).
Compared to the SONG data, the NICMOS data are, in fact, more
sensitive. On average, galaxies with H(AB)=26.1 (M$_B$=--16 at
z$\sim$1, using a Sb spectrum \citealt{kinney96}) in the NICMOS HDFN
are imaged with a S/N$\sim$10 \citep{dickinson00}.  Hence the observed
fraction of barred spirals in the NICMOS HDFN is lower than would have
been observed with a similar magnitude cutoff.  As noted earlier, if
we were to apply a magnitude cutoff similar to \cite{abraham99}, we
would find three barred spirals out of 31 galaxies.  Thus, our finding
of four large bars among 95 galaxies, or three out of 31, indicates
that the fraction of large bars at z $>$ 0.7 seems to be higher than
that found in the local Universe; all of the known selection effects
bias us towards a lower limit to the bar fraction.  We conclude that
the fraction of large barred spiral galaxies is not declining at z $>$
0.7, as previously claimed.

\subsection{Implications of Bars at z$\sim$1}\label{impl}

Observational biases limit us to only identifying the largest bars at
high redshifts.  Though there are significant caveats (small numbers,
difficulty in defining comparable samples at high and low redshifts),
the data suggest that the fraction of large bars at z$\sim$1 may be
higher than that seen in the local Universe.

If the bars we see at z$\sim$1 formed from the bar instability, then
one could infer that cold, rotationally-supported massive disks were
present at least 7 Gyr ago. It is equally, perhaps more, likely that
these bars formed from interactions; however, without additional data
(e.g., observations of gas kinematics in the outer parts of the disk),
it is difficult to distinguish between the two formation scenarios.
If these are merger-induced bars, they indicate that at least some
{\sl large} disks were present at z$\sim$1.  Bars induced by
interactions are likely to be transient phenomena; nevertheless, their
presence at z$\sim$1 suggests that they may have played an important
role in the evolution of galaxy disks, as noted in \S \ref{intro}.  

\section{Conclusions}\label{concl}

At z $<$ 0.7, we find the same number of bars as seen in previous
optical studies.  This is not surprising because bandshifting and
resolution are not a problem at these redshifts.  At z $>$ 0.7, we
identify four barred spirals, doubling the number previously seen.
This result reiterates the known bandshifting effect in identification
of barred spirals.  

Poor spatial resolution can also lead to an underestimation of the bar
fraction.  At 0.8$\micron$, WFPC2 is only marginally capable of
detecting the typical 5 kpc bar beyond z$\sim$0.7. Although the NICMOS
data can compensate for bandshifting, they have even poorer resolution
than WFPC2 and are sensitive only to the largest bars.  The four bars
at z $>$ 0.7 have an average size of 12 kpc.

At z $>$ 0.7 the observed fraction of bars is 4/95 galaxies; all known
selection effects indicate that this fraction is a lower limit.  This
fraction is {\sl higher} than that seen in the local Universe (1/44)
for similarly sized bars.  But there are significant caveats, e.g.,
small numbers, and difficulties in defining comparable samples at high
and low redshifts.  We conclude that there is no significant evidence
for a decrease in barred spirals beyond z$\sim$0.7.

It is difficult to distinguish whether these bars formed in mergers or
from a dynamical instability in the disk.  If it is the latter, then
their presence indicates that cold, massive disks were already present
at z$\sim$1, consistent with star formation history.  If it is the
former, then they indicate that large disks were present at z$\sim$1,
and bars probably played an important role in the evolution of the
galactic disks, and perhaps bulge formation.  Further data are
necessary to fully constrain the cosmological evolution of barred
spirals.

\acknowledgments

We thank the anonymous referee for helpful comments and suggestions
that greatly improved this Letter. We also thank R. Abraham, S. van
den Bergh, D. Block, F. Combes, S. Courteau, R. Ellis, B. Elmegreen,
D. Elmegreen, T. Jarrett, S. Laine, P. Martini, E. Schinnerer, T. Treu,
S. Veilleux, \& C.D. Wilson for useful comments and discussions.  KS
is indebted to Lynne Hillenbrand for her encouragement in applying for
the archival grant.  We thank M. Dickinson and the NICMOS HDF team for
making their reduced data available to us.  Support for Proposal
number No. HST-AR-09552.01-A was provided by NASA through a grant from
the Space Telescope Science Institute, which is operated by the
Association of Universities for Research in Astronomy, Incorporated,
under NASA contract NAS5-26555.  K.S. also acknowledges support from
grant AST-9981546 from the National Science Foundation.


\begin{deluxetable}{rllrrrrrrrrrrr}
\tabletypesize{\scriptsize}
\tablecaption{Barred Galaxies at z $>$ 0.7\label{bartab}}
\tablehead{\colhead{Id} &\colhead{RA} &\colhead{DEC} &\colhead{U$^a$} &\colhead{B$^a$} & \colhead{V$^a$} &\colhead{I$^a$} &\colhead{J$^a$} &\colhead{H$^a$} &\colhead{K$_s$$^a$} &\colhead{phot-z} &\colhead{spec-z}  &\colhead{a} &\colhead{PA}}
\startdata
607 & 12:36:48.78 & 62:13:18.45 & 24.30 & 23.98 & 23.56 & 22.80 & 22.65 & 22.39 & 22.08 & 0.776 & 0.753$^b$ & 1.33 & 104 \\
927 & 12:36:56.78 & 62:13:11.28 & 28.50 & 28.35 & 27.94 & 26.90 & 26.20 & 25.92 & $>$25.56 & 0.904 & ... & 1.02 & 80 \\
1488 & 12:36:46.16 & 62:11:42.13 & 23.44 & 22.80 & 22.28 & 21.30 & 20.49 & 19.88 & 19.31 & 0.904 & 1.016$^c$ & 1.97 & 163 \\
1495 & 12:36:46.87 & 62:11:44.80 & 24.44 & 24.14 & 24.06 & 23.40 & 23.02 & 22.66 & 22.52 & 0.951 & 1.059$^c$ & 1.31 & 1 \\
\enddata
\tablenotetext{a}{AB magnitudes courtesy C. Hanley, M. Dickinson \& the NICMOS HDFN team.}
\tablenotetext{b}{Redshifts compiled on NED by S. Gwyn.  This redshift determined by \citet{lowenthal97}}
\tablenotetext{c}{Redshifts compiled on NED by S. Gwyn. This redshift determined by \citet{cohen96}}
\end{deluxetable}

\end{document}